\documentclass[a4paper,11pt]{article}
\usepackage{jinstpub}
\usepackage{lineno}  
\title{\boldmath Photon reconstruction using the Hough transform in imaging calorimeters}

\author[a,b]{Yang Zhang}
\author[a,b,c,1]{Shengsen Sun,\note{Corresponding author.}}
\author[a,b]{Weizheng Song}
\author[a,d]{Fangyi Guo}
\author[a,b]{Yuanzhan Wang}
\author[a,b,c]{Linghui Wu}
\author[a,b,c]{Yifang Wang}
\affiliation[a]{Institute of High Energy Physics,\\Beijing, 100049, China}
\affiliation[b]{University of Chinese Academy of Sciences,\\Beijing, 100049, China}
\affiliation[c]{High Energy Research Center, Henan Academy of Sciences, \\Zhengzhou, 450046, China}
\affiliation[d]{China Center of Advanced Science and Technology, \\Beijing, 100190, China}

\emailAdd{sunss@ihep.ac.cn}

\abstract
{
Photon reconstruction in calorimeters represents a crucial challenge in particle physics experiments, 
especially in high-density environments where shower overlapping probabilities become significant. 
We present an energy-core-based photon reconstruction method. 
It is achieved through extending the application of the Hough transform to exploit the energy-core structure of photon showers.
The method, validated through simulations of the CEPC crystal electromagnetic calorimeter, achieves a reconstruction efficiency of nearly 100\% for photons with energies exceeding 2 GeV and a separation efficiency approaching 100\% for two 5 GeV photons, when the distance between them reaches the granularity limit of the calorimeter.
This energy-core-based photon reconstruction method, integrated with an energy splitting technique, enhances the performance of photon measurement and provides a promising tool for imaging calorimeters, particularly those requiring high precision in photon detection in complex event topologies with high multiplicity.
}


\arxivnumber{2508.20728}

\begin{document}
\maketitle
\flushbottom

\section{Introduction}
\label{sec:introduction}

Photons, as fundamental constituents of numerous high-energy processes, offer crucial insights into the dynamics and underlying mechanisms of particle interactions.
Hence, accurate photon reconstruction and precise measurement are of great importance in particle physics experiments, enabling precision tests of the Standard Model (SM)~\cite{Glashow:1961tr, Weinberg:1967tq, Salam:1968, GIM:1970} and enhancing sensitivity to new physics.
This task is primarily achieved through the use of electromagnetic calorimeters (ECAL) and their associated reconstruction algorithms. 
However, accurately reconstructing and identifying photons from their electromagnetic (EM) showers in ECAL remains a significant challenge, particularly in environments with spatially close or overlapping particle showers, such as those found in jets. 

The particle flow approach (PFA)~\cite{PFA2001} represents a state-of-the-art methodology for achieving excellent jet energy resolution through the optimal integration of tracking and calorimetric information. 
A critical aspect of the particle flow approach is its ability to accurately associate calorimeter energy depositions with the corresponding reconstructed tracks. 
This imposes stringent requirements on both detector hardware and reconstruction software.
High-granularity is a fundamental requirement for calorimeter design, as it enables the calorimeters to capture the details of shower development through fine spatial segmentation, which is essential for separating nearby showers.
A material with a short radiation length ($X_0$), a small Molière radius ($R_M$), and a high ratio of interaction length to radiation length ($\lambda_I/X_0$) is preferable.
These properties ensure compact EM showers and enable effective longitudinal separation between EM and hadronic showers.
As exemplified by the Silicon-Tungsten sampling ECAL of the International Large Detector (ILD)~\cite{SiW-2008,SiW-2009,SiW-2022,SiW-2020}, where high-resistivity silicon is segmented into small pixels, and tungsten is chosen as absorber material.
These calorimeters provide three-dimensional shower information and precise spacial measurement, enabling excellent jet energy resolution when combined with particle flow algorithms (e.g., PandoraPFA~\cite{PandoraPFA}, Arbor~\cite{Arbor2014,Arbor2018}).
However, its EM energy resolution is inherently limited by the sampling structure.

\begin{figure}[ht]
    \centering
    \includegraphics[width=0.6\textwidth]{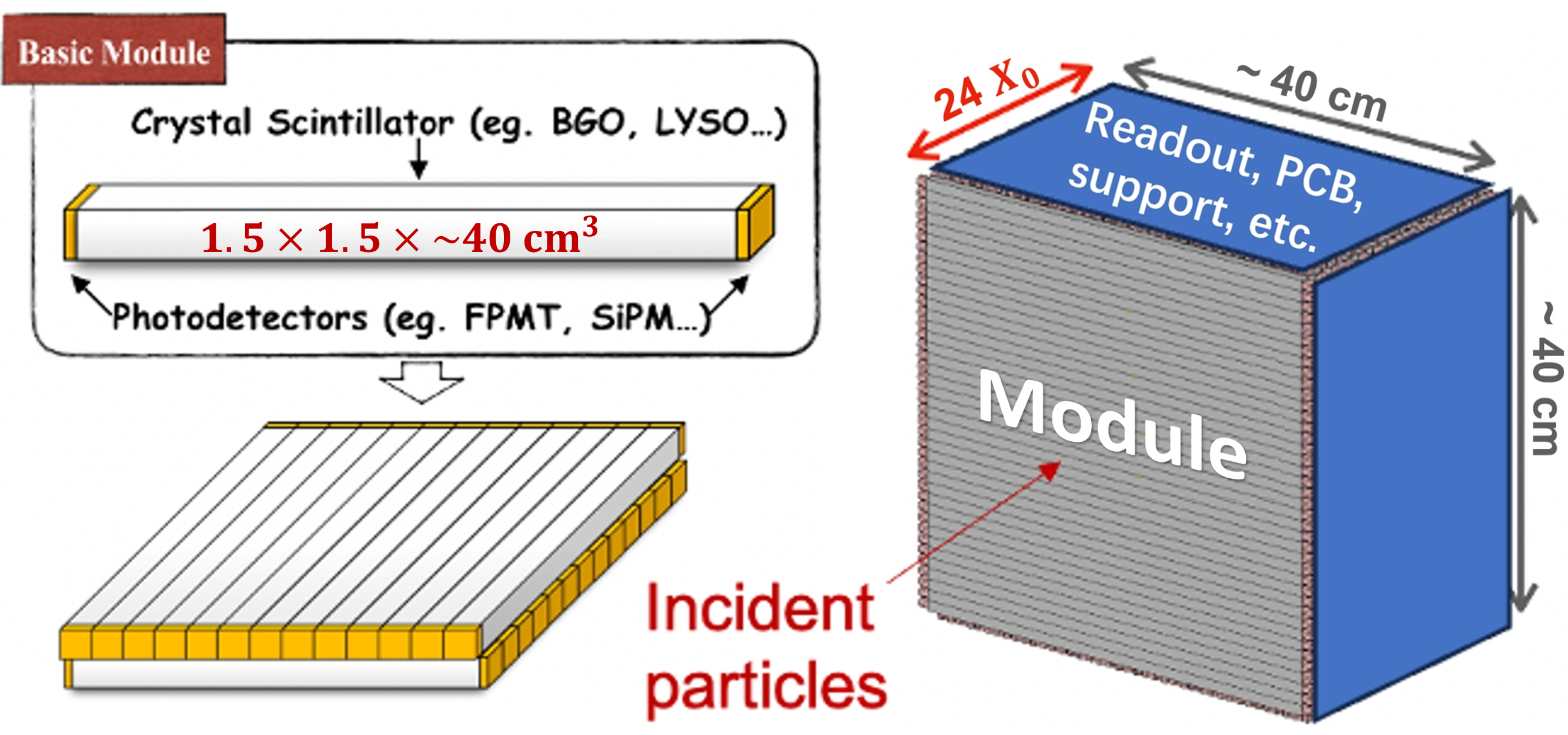}
    \caption{Basic structure of the crystal ECAL. A single crystal bar, with typical dimensions of $1.5\times1.5\times\sim40  \text{cm}^3$, is read out at both ends. Crystals in adjacent layers are oriented perpendicularly to each other. Multiple layers are stacked to form a test module.}
    \label{fig:ECAL_module}
\end{figure}

Homogeneous crystal ECALs provide excellent energy resolution, as they are not subject to the limitations of sampling fluctuations. 
The traditional design of crystal ECALs employs crystal bars with their axes approximately pointed toward the interaction point, which ensures highly granular transverse segmentation. 
However, this geometry inherently lacks longitudinal segmentation. 
Consequently, while these ECALs are powerful instruments for precise EM energy measurement, they cannot perform the detailed three-dimensional shower reconstruction required for Particle Flow Algorithms (PFA).
A viable solution is to implement a crystal ECAL with fine three-dimentional segmentation. 
However, high granularity often necessitates a large number of readout channels, inevitably increasing costs, complexity and power consumption.
To mitigate this, transversely segmented crystal bars offer an effective design alternative~\cite{CrystalECAL}.
The schematic of such a design is shown in Figure.~\ref{fig:ECAL_module}, with a detailed description provided in Section~\ref{sec:CEPC}.
This approach significantly reduces the channel count while simultaneously preserving excellent EM energy resolution and enabling access to three-dimensional shower information.
This design has been adopted by the reference detector of the Circular Electron Position Collider (CEPC)~\cite{CEPCTDRAccelerator,CEPC-Reference-Detector-TDR}.

\begin{figure}[ht]
\centering
\includegraphics[width=0.6\textwidth]{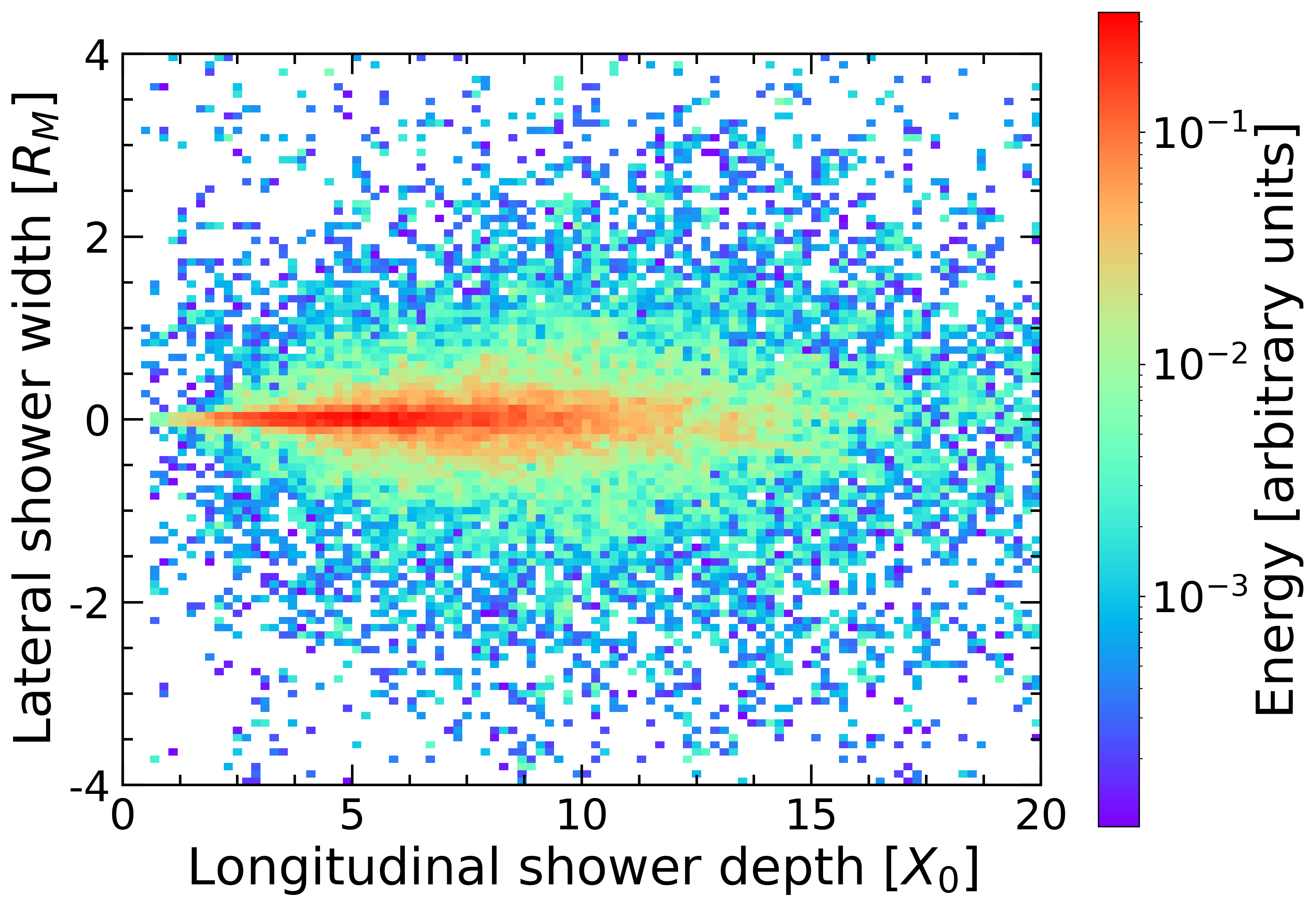}
\caption{ Two-dimensional energy deposition distribution of a 5 GeV photon shower, with the x-axis and y-axis representing longitudinal and lateral directions respectively. A compact energy-core is observed along the direction of shower development.} 
\label{fig:photon_shower}
\end{figure}

As illustrated in Figure~\ref{fig:photon_shower}, at high energy ($E_\gamma>1$~GeV) the profile of the photon shower is characterized by a distinct energy-core along the incident direction, and broadens as the shower develops~\cite{Amaldi:1980uz,Yuda:1969xi}.
This energy-core can be used as a key signature for photon reconstruction and enables effective discrimination against overlapping showers from other particles.
The Hough transform~\cite{HoughTransformation, HoughAlphaRho} is highly effective at identifying specific trajectories, such as lines and circles, even in noisy or high-complexity environments.
It has been a fundamental pattern recognition algorithm for track finding, applicable not only in tracking detectors but also in other detector systems~\cite{SDHCAL-Hough}.

In this paper, a novel photon reconstruction method for imaging calorimeters is proposed through extending the application of the Hough transform to exploit the energy-core structure of photon showers.
For the first time, this technique is extended to the specific challenge of photon reconstruction. 
The integration of this method and energy splitting technique has been implemented and validated using a simulated crystal ECAL for the CEPC.

\section{Crystal ECAL design and its simulation}
\label{sec:CEPC}

Long bar crystals, made of bismuth germanate (BGO), are the basic detection units of the crystal ECAL of the CEPC reference detector~\cite{CEPC-Reference-Detector-TDR}.
Each crystal is read out by silicon photomultipliers (SiPMs) at both ends, enabling precise measurement of energy and time.
The crystals in adjacent layers are arranged perpendicularly to each other.
Multiple layers of crystals form a module. 
Figure~\ref{fig:ECAL_module} illustrates the basic structures of a crystal, two adjacent layers and a test module.
The transverse size of each crystal is 15 mm, corresponding to approximately 0.67 Molière radius of BGO, which is 22.3 mm.
This crisscross arrangement indicates that this ECAL does not function as a traditional calorimeter with crystals always toward the incident particles.

\begin{figure}[htbp]
    \centering
    \includegraphics[width=.4\textwidth]{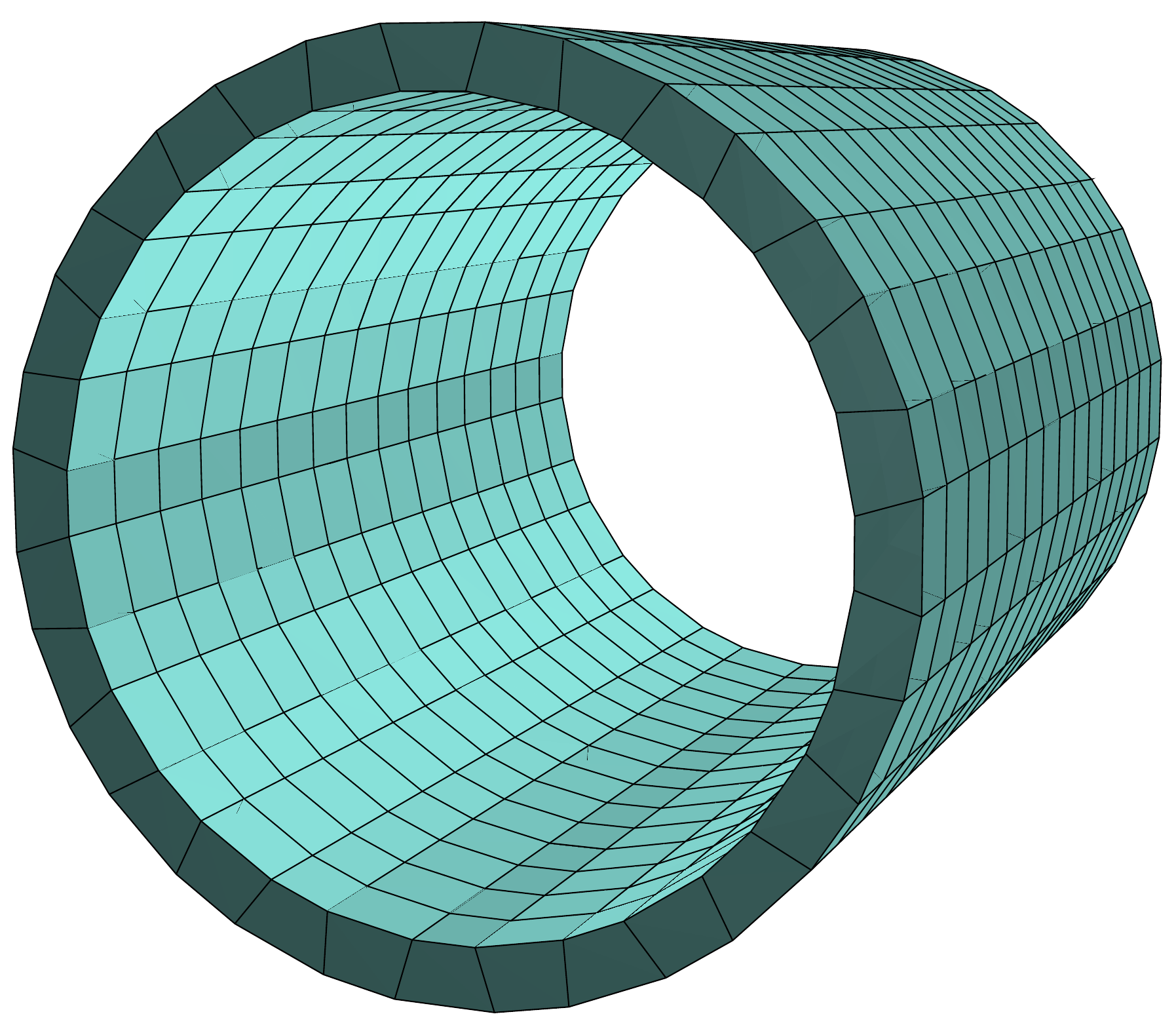}
    \qquad
    \includegraphics[width=.4\textwidth]{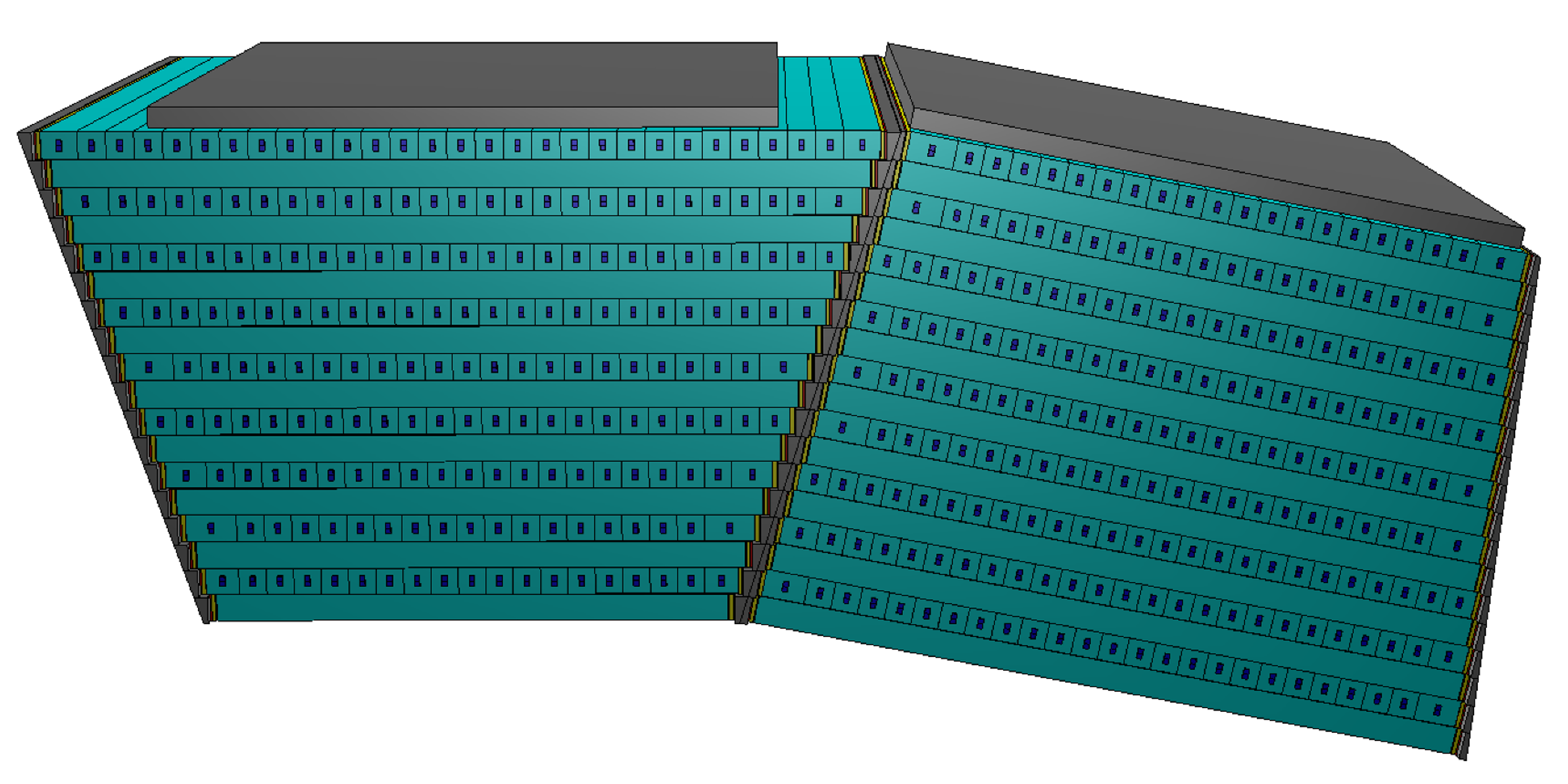}
    \caption{Global view of the barrel ECAL (left) and schematic of two adjacent trapezoidal modules (right).\label{fig:ecal_structure}}
\end{figure}

The global view of the barrel ECAL~\cite{Qi:2025dvo} of the CEPC reference detector is shown in the left plot of Figure~\ref{fig:ecal_structure}.
Each module is longitudinally segmented into 18 layers, and has a total thickness of about 24 $X_0$. 
The modules are designed with trapezoidal shapes, as shown in right plot of Figure~\ref{fig:ecal_structure}, to prevent inter-module gaps from pointing toward the interaction point.

The energy-core based photon reconstruction method is developed under the CEPC software (CEPCSW)~\cite{CEPCSW}. CEPCSW leverages multiple established high-energy physics software packages, enabling comprehensive detector design optimization and physics performance studies.
This work is based on full simulations of the reference CEPC detector in CECPSW framework.
The material and geometry descriptions of the ECAL are implemented via DD4hep~\cite{frank_markus_2018_1464634}, incorporating both the sensitive volumes and dead regions, including BGO crystal, reflective wrappers, readout electronics, cooling system and mechanical support structures.
The GEANT4 version 11.2.0~\cite{Geant4-2003, Geant4-2006, Geant4-2016} provides a well proofed description of the energy deposition in the crystals by the EM showers.
The digitization process models the scintillation photon statistics, SiPM response, and readout electronics, with parameters from dedicated R\&D works~\cite{CEPC-Reference-Detector-TDR}.
To ensure a conservative performance estimates, Monte Carlo (MC) truth information is strictly excluded in the development of photon reconstruction and energy splitting algorithms.

\section{Photon reconstruction with Hough transform}
\label{sec:Hough}

The crystals in the odd and even layers form two projection planes, both perpendicular to the longitudinal axis of the crystals. 
The photon reconstruction process using the Hough transform is performed through the following steps in both projection planes.
The results are subsequently combined.

\subsection{Clustering and local maximum selection}
\begin{figure}[ht]
\centering
\includegraphics[width=0.6\textwidth]{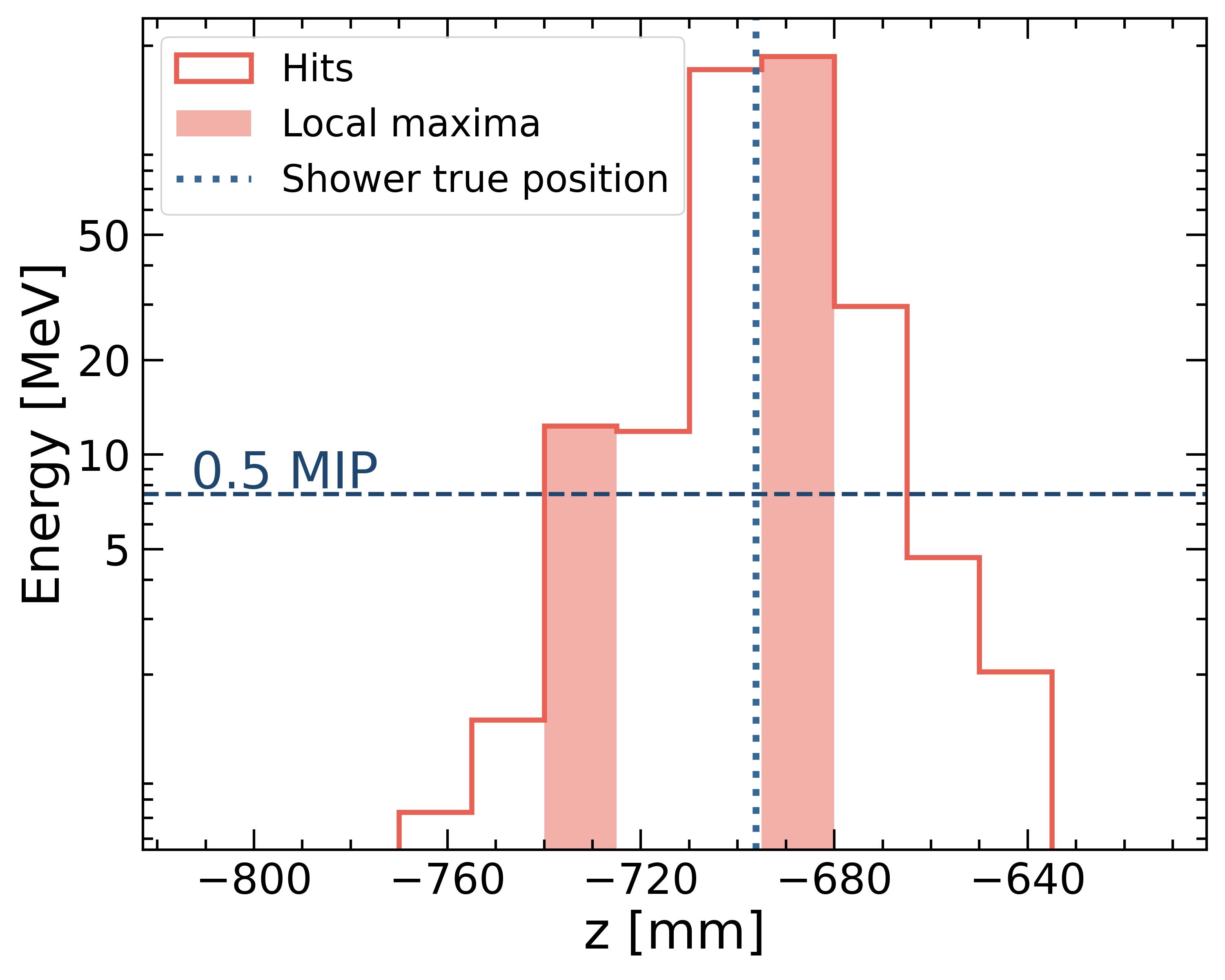}
\caption{Example of the energy deposition distribution of crystals in one layer, for a photon of 5 GeV, where each bin corresponds to a crystal. Crystals satisfying the local maximum condition are highlighted. The vertical dotted line indicates the true position of the shower}
\label{fig:local_max}
\end{figure}

A clustering algorithm aggregates adjacent crystals with energy depositions above a defined threshold.
The resulting collection is defined as a cluster, representing either a single particle's shower or the overlapping showers of multiple particles. 
A local maximum is defined as a crystal whose energy deposition exceeds that of its two immediate neighbors in the same layer and must also surpass a specific value --- chosen here as half the energy deposition of a minimum ionizing particle (MIP).
For the example shown in Figure~\ref{fig:local_max}, two local maxima are found in the energy deposition distribution of crystals in one layer.
The higher one is located in the central region of the shower, corresponding to the position of the shower energy-core, while the lower one arises from stochastic fluctuations.
  
\begin{figure}[ht]
  \centering
  \includegraphics[width=0.6\textwidth]{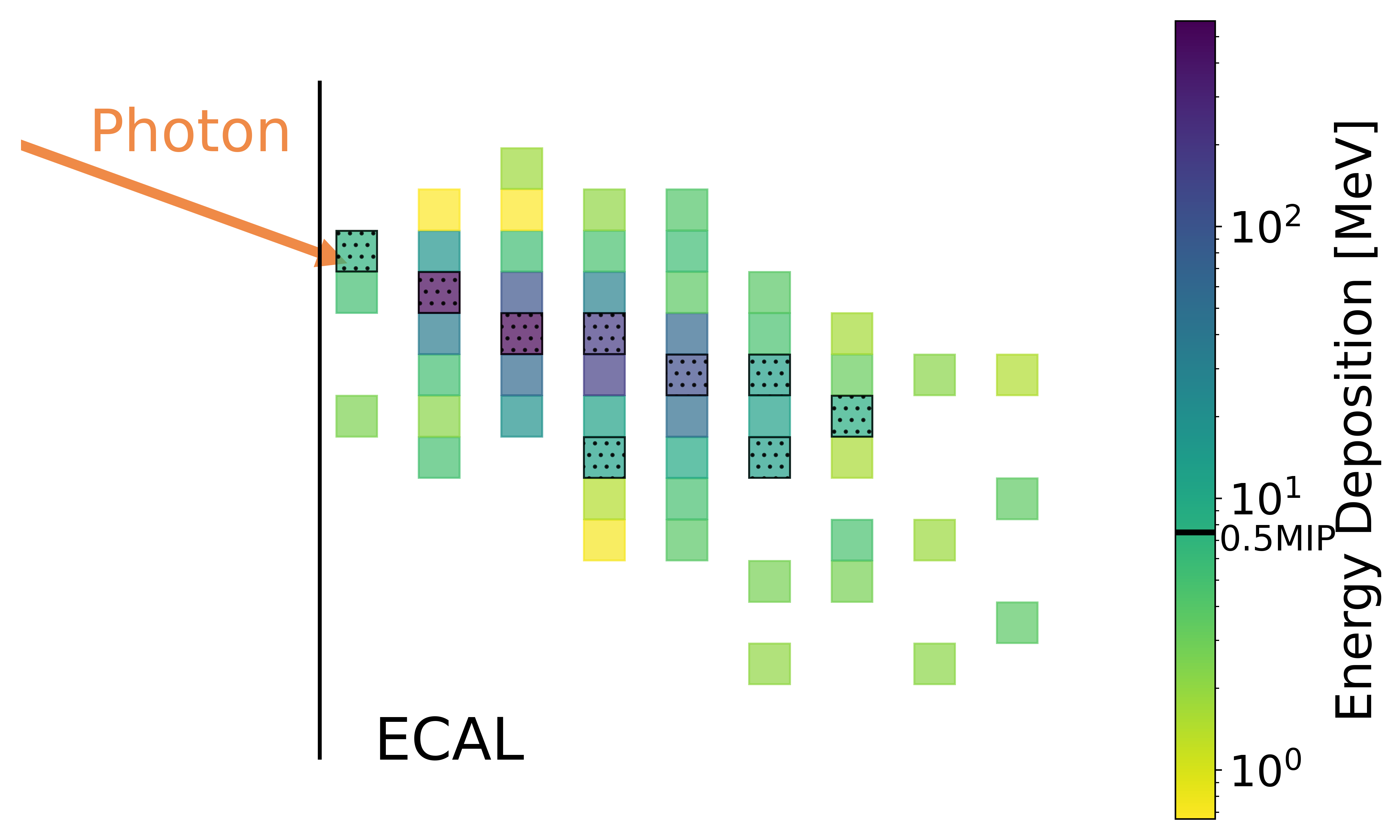}
  \caption{Spatial distribution of crystals for a 5 GeV photon, with local maxima highlighted. The color scale represents the energy deposition, with the energy threshold (0.5 MIP) of local maxima indicated on the color bar. Local maxima are marked with dot patterns.}
  \label{fig:local_max_comparison}
\end{figure}

The spatial distributions of crystals with energy deposition and local maxima in one of the projection planes are shown in Figure~\ref{fig:local_max_comparison}, for one shower of 5 GeV, with the incident direction of the photon indicated by an arrow.
The majority of local maxima are distributed around the extension of the photon incident direction, consistent with the expected concentration of energy deposition along the photon shower axis.
A small fraction of local maxima deviate from this direction, due to fluctuations in energy deposition.

\subsection{Hough transform for photon reconstruction}
\begin{figure}[ht]
  \centering
  \includegraphics[width=0.6\textwidth]{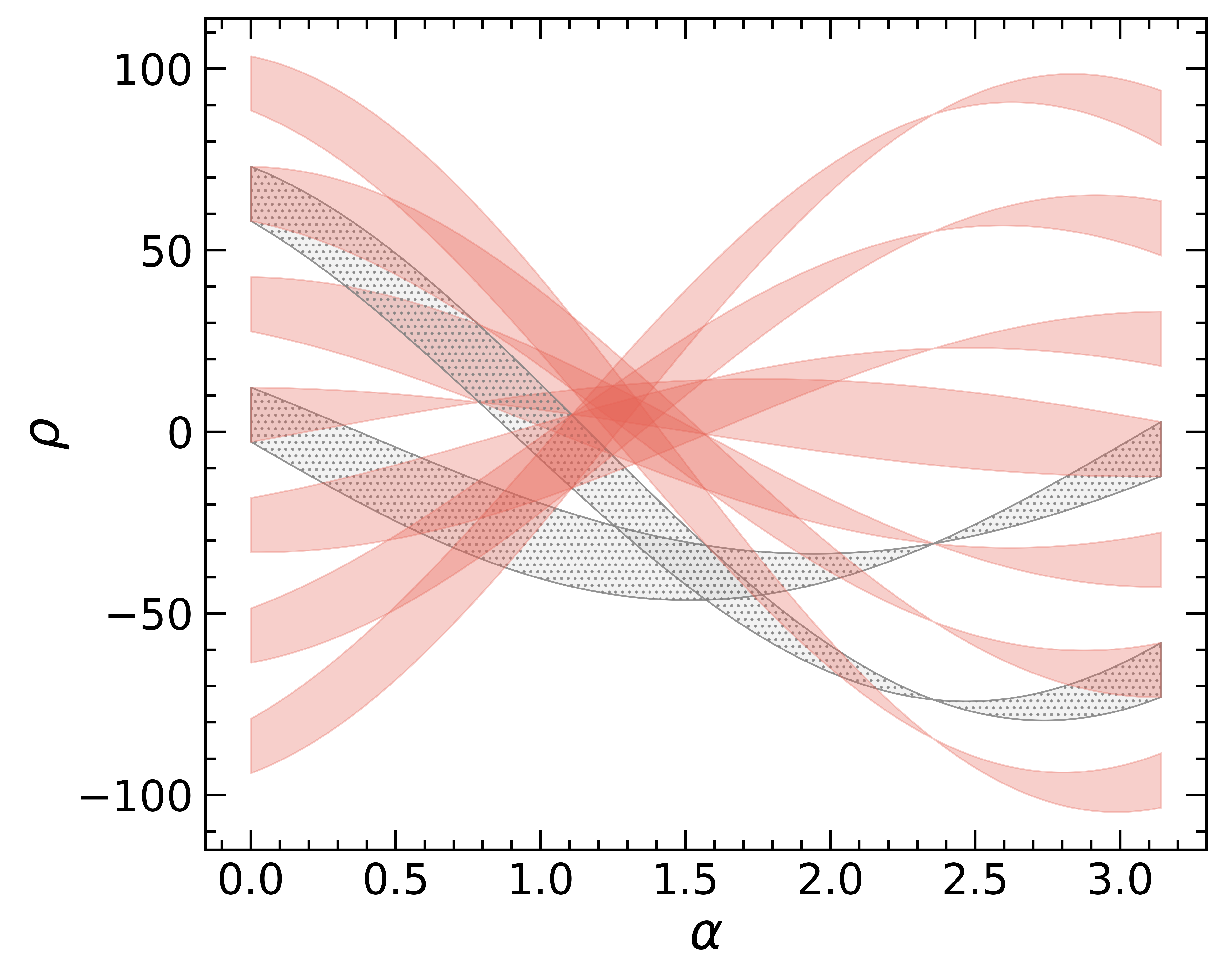}
  \caption{Distribution of band regions for a 5 GeV photon shower. Parameters $\rho$ and $\alpha$ are the distance and angle, respectively, of the Hough-transformed line in the image space. The Hough space uses a bin size of approximately half the crystal transverse dimension, chosen to balance reconstruction performance with computational cost. Each band corresponds to a local maximum transformed into Hough space. The red bands originate from local maxima on the energy-core, intersecting in a common region (around $\alpha=1.3$, $\rho=0$) that defines the Hough axis of the shower. The gray, hatched bands arise from fluctuation-induced local maxima and do not participate in forming a consistent axis.}
  \label{fig:hough_band}
\end{figure}

\begin{figure}[ht]
  \centering
  \includegraphics[width=0.6\textwidth]{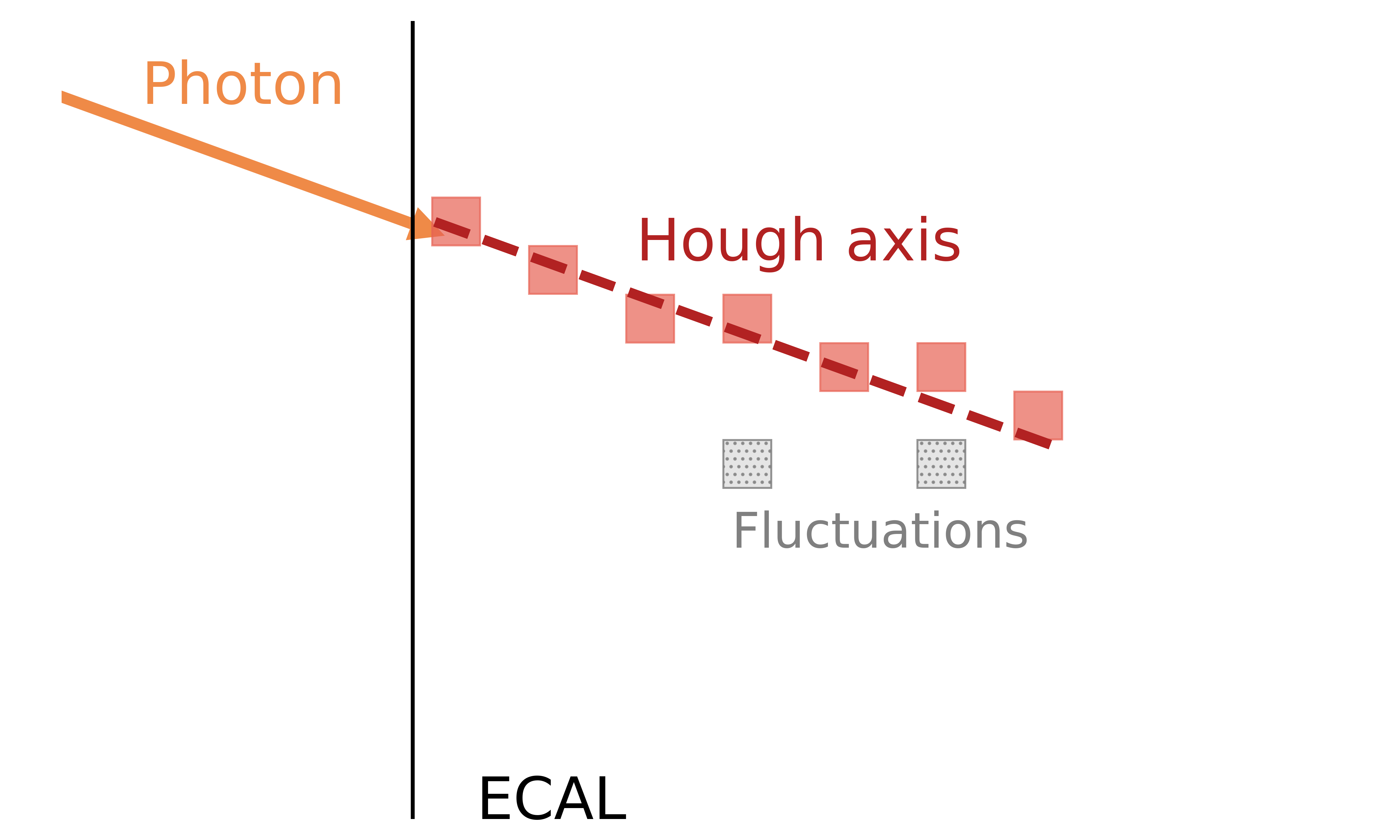}
  \caption{Identification of the energy-core for a 5 GeV photon shower represented as a Hough axis. The dashed line marks the reconstructed Hough axis, which aligns with the incident direction of the photon (orange arrow). The red local maxima lying on the Hough axis constitute the identified energy-core of the shower. The gray, hatched local maxima arise from energy fluctuations and are rejected by the algorithm as they do not belong to the consistent axis.}
  \label{fig:hough_axis}
\end{figure}

We develop a generalized Hough transform, extending the conventional point-like hit assumption used in charged particle tracking.
In the generalized approach, each local maximum in the image space is represented as a square whose position and dimensions correspond to the location and the transverse geometry of the crystal.
Consequently, a local maximum in the image space is transformed into a band-shaped region in the Hough space, 
whose width corresponds to the projection width of the crystal as seen from the interaction region, 
in contrast to the single curve generated by the traditional approach. 
Each point within this band corresponds to a straight line in the image space that intersects the local maximum. 
When multiple local maxima are considered, their corresponding bands in the Hough space may overlap, as shown in Figure~\ref{fig:hough_band}. 
The intersection of these bands defines a set of lines in the image space that simultaneously pass through all local maxima, thereby indicating their collinearity.
These collinear local maxima are grouped into a Hough axis, which represents the energy-core of a photon shower, as illustrated by the dashed line in Figure~\ref{fig:hough_axis}.
Each local maximum on the Hough axis is defined as the "seed" crystal of the layer.
The Hough axis is required to contain at least three consecutive local maxima and must point toward the interaction point to ensure the robustness of the algorithm.
This constraint effectively suppresses the reconstruction of "fake" photon showers caused by energy fluctuations.
To handle overlapping showers, a local maximum can be shared between two adjacent Hough axes, provided that the energy deposition in the shared crystals does not exceed half of the total energy associated with either axis. This provision enhances the algorithm's ability to resolve closely spaced photons.
Consequently, a combination of identified Hough axes from each projection plane that meet the matching criteria is interpreted as a photon candidate.

\subsection{Energy splitting for overlapping showers}
\begin{figure}[ht]
    \centering
    \includegraphics[width=0.6\textwidth]{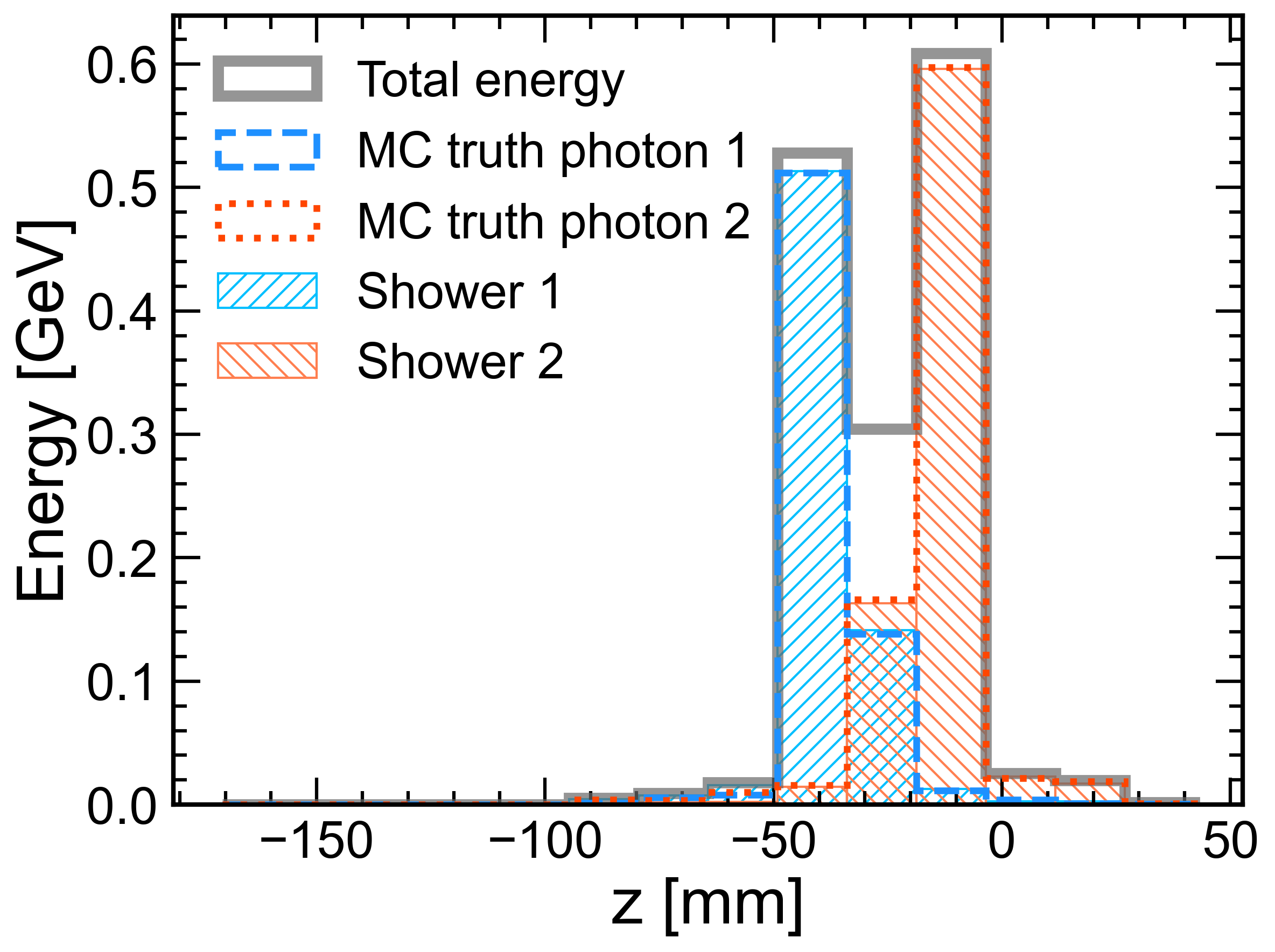}
    \caption{Illustration of energy splitting for two overlapping photon showers. The gray open histogram shows the total energy deposition of two overlapping photons. The reconstructed energy profiles assigned to each photon by the splitting algorithm are shown as hatched histograms. The true energy profiles from the MC simulation are overlaid for comparison.}
    \label{fig:energy_splitting}
\end{figure}

When two or more photons are sufficiently close in space, their EM showers overlap.
The energy deposition from different showers in each crystal should be properly allocated to the corresponding photon through an energy splitting algorithm.
Based on  the lateral development of EM showers, the energy splitting algorithm employ the sum of two exponential functions~\cite{Akopdzhanov:1976pr} to simultaneously describe the narrow core and its surrounding halo.
For a photon $\mu$ with energy $E^{\text{seed}}$ in "seed" crystal, the expected energy deposition in crystal $i$ is given by:
\begin{equation}
    E_{i\mu}^{\text{exp}} = E^{\text{seed}} \times f(r_i),
\end{equation}
  where 
\begin{equation}
    f(r_i) = p_0 \exp{\left(-\frac{p_2r_i}{R_M}\right)} + p_1 \exp{\left(-\frac{p_3r_i}{R_M}\right)}.
\end{equation}
Here, $f(r_i)$ describes the lateral profile of the energy deposition in a photon shower, $p_0$ and $p_1$ are fractional parameters, $p_2$ and $p_3$ are utilized to describe the lateral width of the shower, $R_M$ is the Molière radius of the crystal, and $r_i$ is the distance from the crystal $i$ to the reconstructed shower position, with the center of the "seed" crystal serving as the initial estimation.
Let $E_i$ be the total energy deposited in the crystal $i$,
the expected energy assigned to photon $\mu$ in crystal $i$ is then calculated as:
\begin{equation}
    E_{i\mu}=\frac{E_{i\mu}^{\text{exp}}}{\sum_\mu E_{i\mu}^{\text{exp}}} \times E_i.
\end{equation}
The energy deposition in each crystal is reallocated to the individual photons according to their expected fractional contributions.
$\vec{x_\mu}$ is the reconstructed shower position of photon $\mu$, which is calculated using a center-of-gravity method:
\begin{equation}
    \vec{x_{\mu}}=\frac{\sum_i E_{i\mu}{\cdot}{\vec{x_i}}}{\sum_i E_{i\mu}},
\end{equation}
where $\vec{x_i}$ is the center of the crystal $i$.
This procedure converges to a stable solution after several iterations.
Figure~\ref{fig:energy_splitting} shows, as an example, the energy distributions of two overlapping photons.
Each bin in the histograms represents the energy deposition in a single crystal. 
The gray open histogram outlines the total energy deposition of two overlapping photons. The energy splitting algorithm successfully disentangles their energy deposition, allocating the energy into two distinct showers, represented by blue and red hatched histograms.
A close agreement is observed between these split energies and the true profiles from the MC simulation (dashed and dotted histograms), demonstrating the effectiveness of the method.
The slight difference between the split energy and that of MC truth is caused by the fluctuation of the energy deposition.

\section{Performance}
\label{sec:Performance}
A performance study of the energy-core-based photon reconstruction and energy splitting algorithms is presented using MC simulation of single-photon and two-photon events.
The study focuses on quantifying the reconstruction efficiency of photon-induced EM showers and the separation efficiency of two spatially overlapping EM showers.

\subsection{Single photon reconstruction}
\label{sec:single-photon}
\begin{figure}[ht]
    \centering
    \includegraphics[width=0.6\textwidth]{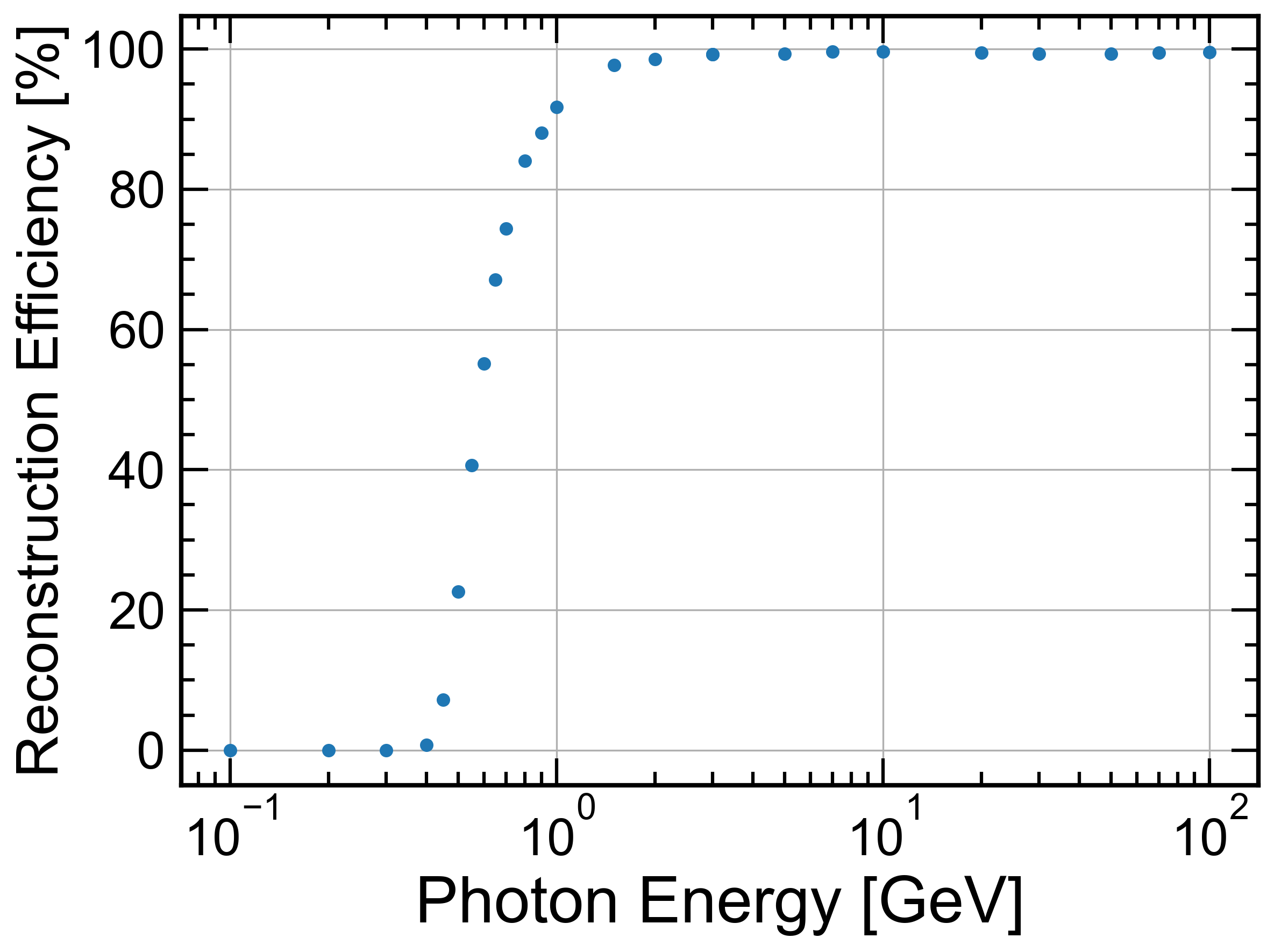}
    \caption{Single-photon reconstruction efficiency as a function of incident photon energy in the range of 0.1 to 100 GeV.}
    \label{fig:single_photon_efficiency_15mm}
\end{figure}

The single-photon reconstruction efficiency is determined using MC samples of single-photon events spanning an energy range from 0.1 GeV to 100 GeV. 
The photons are generated with an uniform distribution in azimuthal angle $\phi$ range from $0^\circ$ to $360^\circ$ and polar angle $\theta$ range from $40^\circ$ to $140^\circ$. 
This $\theta$ range ensures full containment of the photons within the barrel ECAL's acceptance and minimal energy leakage at the boundaries.
The reconstruction efficiency is defined as the ratio of the number of good photon candidates to the total number of unconverted photons in the sample.
A good photon candidate must satisfy the following two criteria: a reconstructed cluster energy containing at least 75\% of the total deposited energy; an angular difference between the reconstructed direction and the MC truth incident direction of less than $0.25^\circ$.
This maximum angle difference corresponds to a transverse distance slightly larger than half the width of a single crystal.
As shown in Figure~\ref{fig:single_photon_efficiency_15mm}, the reconstruction efficiency reaches nearly 100\% for photons above 2 GeV; 50\% are achieved for 0.6 GeV photons.
For photon energies below 2 GeV, photons traverse insufficient longitudinal layers to be effectively reconstructed by the Hough transform, resulting in a significant reduction in reconstruction efficiency.

\subsection{Two-photon separation}
The two-photon separation efficiency is investigated as a function of the distance between the incident positions of two photons, which is used to evaluate the performance of in resolving overlapping EM showers. 
A MC sample of two photons is generated, with each photon having an energy of 5 GeV.
The $\phi$ angles of two photons are fixed at $0^\circ$, while each of their $\theta$ angles is uniformly distributed between $88.4^\circ$ and $91.6^\circ$, thus ensuring that both photons are incident on the central region of the same module.
This corresponding to a distance range of 0 to 100 mm on the ECAL surface.
The definition of a good photon candidate is the same as described in Section~\ref{sec:single-photon}.
The separation efficiency is defined as the ratio of the number of events with two photon candidates to the number of generated events without photon conversion.
  
\begin{figure}[ht]
    \centering
    \includegraphics[width=0.6\textwidth]{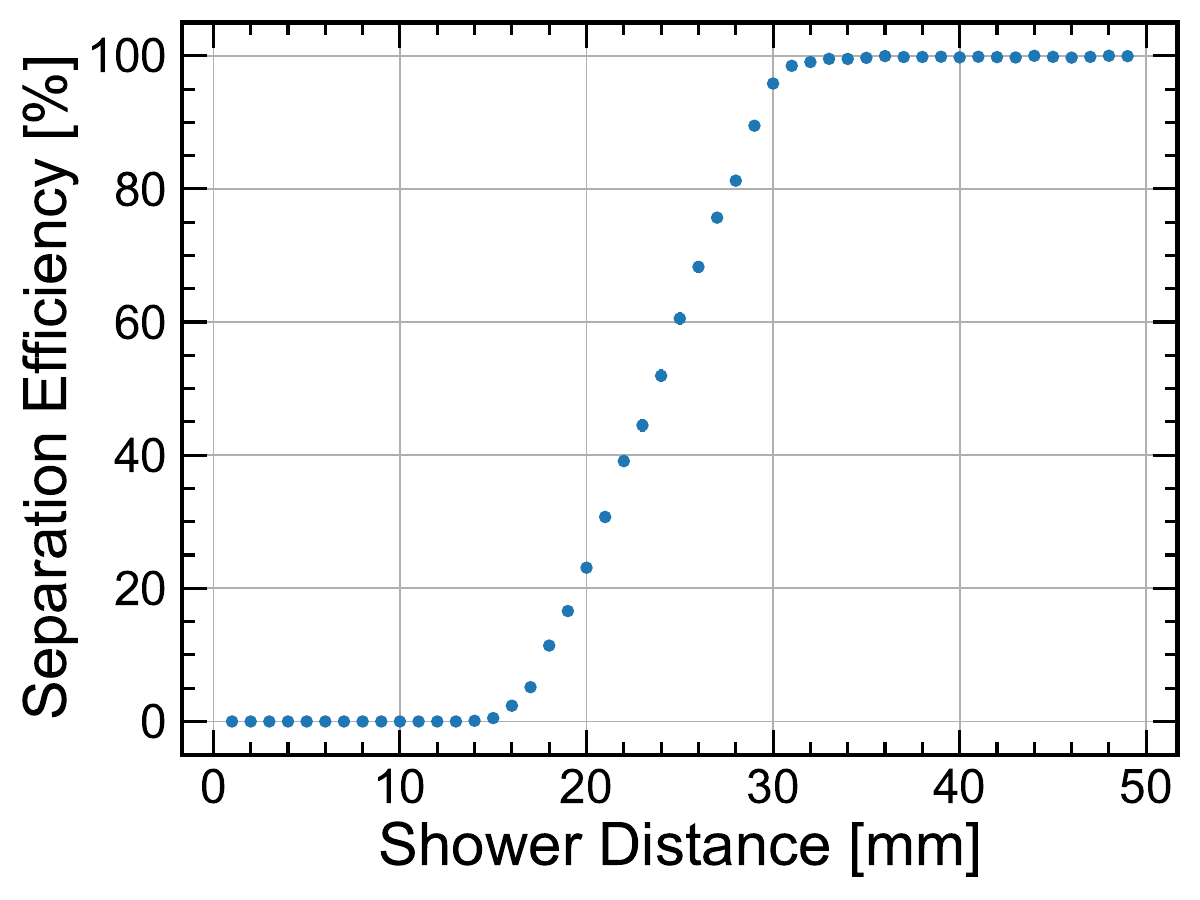}
    \caption{Two-photon separation efficiency as a function of the distance between showers.}
    \label{fig:di_photon_efficiency_15mm}
\end{figure}
As shown in Figure~\ref{fig:di_photon_efficiency_15mm}, the separation efficiency increases from 0\% to approximate 100\% when the distance between the two photons exceeds 15 mm and 30 mm, respectively.
Given the cross-sectional dimensions of the crystals 15 mm$\times$15 mm, the spatial separation of 15 mm and 30 mm between the two photons corresponds to the width of one and two crystals, equivalent to 0.67 and 1.35 Molière radii.
This result demonstrates the excellent photon separation capability of the energy-core-based reconstruction method, and the limitation in separation efficiency is dictated by the granularity of the detector, rather than the scale of the Molière radius or the degree of shower overlap, when the granularity is comparable to Molière radius.

\section{Summary}
\label{sec:Conclusions}
Based on the underlying mechanism of EM shower development, a highly localized energy-core is intrinsically formed along the central axis of the shower.
Leveraging this universal feature of EM showers, an innovative energy-core-based photon reconstruction method for imaging calorimeters has been developed, implementing the application of the Hough transform for photon reconstruction for the first time.
This method, when combined with energy splitting technique, achieves high single-photon reconstruction efficiency and two-photon separation capability, as validated by the simulations of the CEPC crystal ECAL.
Its integration into particle flow algorithms is expected to  enhance their overall performance.
Furthermore, the energy-core is an intrinsic property of electromagnetic shower development, independent of the specific ECAL technology employed.
This grants the method ready generalizability across a wide range of imaging calorimeters.
  
By substantially reducing the reliance on a well defined boundaries of photon shower profiles, this integrated approach alleviates the stringent requirements for both fine granularity and compactness of EM showers, such as short radiation length and Molière radius of the material, in particle flow ECAL design.
This advancement enables greater flexibility in selecting detector technologies, thus facilitating an optimal balance among cost, fabrication complexity, and performance.

The success of the energy-core-based photon reconstruction method also opens up new possibilities for applying advanced pattern recognition algorithms.
Notably, established techniques, already demonstrated to be effective in track-finding, could also be adopted to enhance photon reconstruction efficiency in calorimeters.

\acknowledgments
The authors would like to thank Tianchi Zhao of Washington University for helpful discussions on topics related to this work.

This work is supported in part by the National Natural Science Foundation of China under grant No. 12342502.

\section*{Code Availability Statement}
The algorithm developed in this study is available upon reasonable request from the corresponding author.

% Bibliography
%% [A] Recommended: using JHEP.bst file
\bibliographystyle{JHEP}
\bibliography{biblio.bib}

@article{Glashow:1961tr,
    author = "Glashow, S. L.",
    title = "{Partial Symmetries of Weak Interactions}",
    doi = "10.1016/0029-5582(61)90469-2",
    journal = "Nucl. Phys.",
    volume = "22",
    pages = "579--588",
    year = "1961"
}

@article{Weinberg:1967tq,
    author = "Weinberg, Steven",
    title = "{A Model of Leptons}",
    doi = "10.1103/PhysRevLett.19.1264",
    journal = "Phys. Rev. Lett.",
    volume = "19",
    pages = "1264--1266",
    year = "1967"
}

@inbook{Salam:1968,
    author = {Abdus Salam},
    title = {Weak and electromagnetic interactions},
    booktitle = {Selected Papers of Abdus Salam},
    chapter = {},
    pages = {244-254},
    doi = {10.1142/9789812795915_0034},
    year="1968"
}

@article{GIM:1970,
  title = {Weak Interactions with Lepton-Hadron Symmetry},
  author = {Glashow, S. L. and Iliopoulos, J. and Maiani, L.},
  journal = {Phys. Rev. D},
  volume = {2},
  issue = {7},
  pages = {1285--1292},
  numpages = {0},
  year = {1970},
  month = {Oct},
  publisher = {American Physical Society},
  doi = {10.1103/PhysRevD.2.1285},
  url = {https://link.aps.org/doi/10.1103/PhysRevD.2.1285}
}

@article{PFA2001,
    author = "Brient, Jean-Claude and Videau, Henri",
    editor = "Graf, Norman",
    title = "{The Calorimetry at the future e+ e- linear collider}",
    eprint = "hep-ex/0202004",
    archivePrefix = "arXiv",
    reportNumber = "SNOWMASS-2001-E3047",
    journal = "eConf",
    volume = "C010630",
    pages = "E3047",
    year = "2001"
}

@article{CrystalECAL,
    author = {Qi, Baohua and Liu, Yong},
    title = {Research and Development of a Novel High Granularity Crystal Electromagnetic Calorimeter},
    journal = {Instruments},
    volume = {6},
    year = {2022},
    url = {https://www.mdpi.com/2410-390X/6/3/40},
    issn = {2410-390X},
    doi = {10.3390/instruments6030040}
}

@article{CEPCTDRAccelerator,
    author={{The CEPC Study Group}},
    title = {{CEPC Technical Design Report: Accelerator}},
    eprint = "2312.14363",
    archivePrefix = "arXiv",
    primaryClass = "physics.acc-ph",
    reportNumber = "IHEP-CEPC-DR-2023-01, IHEP-AC-2023-01",
    doi = "10.1007/s41605-024-00463-y",
    journal = "Radiat. Detect. Technol. Methods",
    volume = "8",
    number = "1",
    pages = "1--1105",
    year = "2024"
}

@misc{frank_markus_2018_1464634,
author       = {Frank, Markus and Gaede, Frank and Petric, Marko and Sailer, Andre},
title        = {AIDASoft/DD4hep},
month        = oct,
year         = 2018,
note         = {webpage: http://dd4hep.cern.ch/},
doi          = {10.5281/zenodo.592244},
url          = {https://doi.org/10.5281/zenodo.592244}
}

@article{Geant4-2003,
title = {Geant4—a simulation toolkit},
journal = {Nuclear Instruments and Methods in Physics Research Section A: Accelerators, Spectrometers, Detectors and Associated Equipment},
volume = {506},
number = {3},
pages = {250-303},
year = {2003},
issn = {0168-9002},
doi = {https://doi.org/10.1016/S0168-9002(03)01368-8},
url = {https://www.sciencedirect.com/science/article/pii/S0168900203013688},
author = {S. Agostinelli and J. Allison and K. Amako and J. Apostolakis and H. Araujo and P. Arce and M. Asai and D. Axen and S. Banerjee and G. Barrand and F. Behner and L. Bellagamba and J. Boudreau and L. Broglia and A. Brunengo and H. Burkhardt and S. Chauvie and J. Chuma and R. Chytracek and G. Cooperman and G. Cosmo and P. Degtyarenko and A. Dell'Acqua and G. Depaola and D. Dietrich and R. Enami and A. Feliciello and C. Ferguson and H. Fesefeldt and G. Folger and F. Foppiano and A. Forti and S. Garelli and S. Giani and R. Giannitrapani and D. Gibin and J.J. {Gómez Cadenas} and I. González and G. {Gracia Abril} and G. Greeniaus and W. Greiner and V. Grichine and A. Grossheim and S. Guatelli and P. Gumplinger and R. Hamatsu and K. Hashimoto and H. Hasui and A. Heikkinen and A. Howard and V. Ivanchenko and A. Johnson and F.W. Jones and J. Kallenbach and N. Kanaya and M. Kawabata and Y. Kawabata and M. Kawaguti and S. Kelner and P. Kent and A. Kimura and T. Kodama and R. Kokoulin and M. Kossov and H. Kurashige and E. Lamanna and T. Lampén and V. Lara and V. Lefebure and F. Lei and M. Liendl and W. Lockman and F. Longo and S. Magni and M. Maire and E. Medernach and K. Minamimoto and P. {Mora de Freitas} and Y. Morita and K. Murakami and M. Nagamatu and R. Nartallo and P. Nieminen and T. Nishimura and K. Ohtsubo and M. Okamura and S. O'Neale and Y. Oohata and K. Paech and J. Perl and A. Pfeiffer and M.G. Pia and F. Ranjard and A. Rybin and S. Sadilov and E. {Di Salvo} and G. Santin and T. Sasaki and N. Savvas and Y. Sawada and S. Scherer and S. Sei and V. Sirotenko and D. Smith and N. Starkov and H. Stoecker and J. Sulkimo and M. Takahata and S. Tanaka and E. Tcherniaev and E. {Safai Tehrani} and M. Tropeano and P. Truscott and H. Uno and L. Urban and P. Urban and M. Verderi and A. Walkden and W. Wander and H. Weber and J.P. Wellisch and T. Wenaus and D.C. Williams and D. Wright and T. Yamada and H. Yoshida and D. Zschiesche}
}

@ARTICLE{Geant4-2006,
  author={Allison, J. and Amako, K. and Apostolakis, J. and Araujo, H. and Arce Dubois, P. and Asai, M. and Barrand, G. and Capra, R. and Chauvie, S. and Chytracek, R. and Cirrone, G.A.P. and Cooperman, G. and Cosmo, G. and Cuttone, G. and Daquino, G.G. and Donszelmann, M. and Dressel, M. and Folger, G. and Foppiano, F. and Generowicz, J. and Grichine, V. and Guatelli, S. and Gumplinger, P. and Heikkinen, A. and Hrivnacova, I. and Howard, A. and Incerti, S. and Ivanchenko, V. and Johnson, T. and Jones, F. and Koi, T. and Kokoulin, R. and Kossov, M. and Kurashige, H. and Lara, V. and Larsson, S. and Lei, F. and Link, O. and Longo, F. and Maire, M. and Mantero, A. and Mascialino, B. and McLaren, I. and Mendez Lorenzo, P. and Minamimoto, K. and Murakami, K. and Nieminen, P. and Pandola, L. and Parlati, S. and Peralta, L. and Perl, J. and Pfeiffer, A. and Pia, M.G. and Ribon, A. and Rodrigues, P. and Russo, G. and Sadilov, S. and Santin, G. and Sasaki, T. and Smith, D. and Starkov, N. and Tanaka, S. and Tcherniaev, E. and Tome, B. and Trindade, A. and Truscott, P. and Urban, L. and Verderi, M. and Walkden, A. and Wellisch, J.P. and Williams, D.C. and Wright, D. and Yoshida, H.},
  journal={IEEE Transactions on Nuclear Science}, 
  title={Geant4 developments and applications}, 
  year={2006},
  volume={53},
  number={1},
  pages={270-278},
  doi={10.1109/TNS.2006.869826}}

@article{Geant4-2016,
    title = {Recent developments in Geant4},
    journal = {Nuclear Instruments and Methods in Physics Research Section A: Accelerators, Spectrometers, Detectors and Associated Equipment},
    volume = {835},
    pages = {186-225},
    year = {2016},
    issn = {0168-9002},
    doi = {https://doi.org/10.1016/j.nima.2016.06.125},
    url = {https://www.sciencedirect.com/science/article/pii/S0168900216306957},
    author = {J. Allison and K. Amako and J. Apostolakis and P. Arce and M. Asai and T. Aso and E. Bagli and A. Bagulya and S. Banerjee and G. Barrand and B.R. Beck and A.G. Bogdanov and D. Brandt and J.M.C. Brown and H. Burkhardt and Ph. Canal and D. Cano-Ott and S. Chauvie and K. Cho and G.A.P. Cirrone and G. Cooperman and M.A. Cortés-Giraldo and G. Cosmo and G. Cuttone and G. Depaola and L. Desorgher and X. Dong and A. Dotti and V.D. Elvira and G. Folger and Z. Francis and A. Galoyan and L. Garnier and M. Gayer and K.L. Genser and V.M. Grichine and S. Guatelli and P. Guèye and P. Gumplinger and A.S. Howard and I. Hřivnáčová and S. Hwang and S. Incerti and A. Ivanchenko and V.N. Ivanchenko and F.W. Jones and S.Y. Jun and P. Kaitaniemi and N. Karakatsanis and M. Karamitros and M. Kelsey and A. Kimura and T. Koi and H. Kurashige and A. Lechner and S.B. Lee and F. Longo and M. Maire and D. Mancusi and A. Mantero and E. Mendoza and B. Morgan and K. Murakami and T. Nikitina and L. Pandola and P. Paprocki and J. Perl and I. Petrović and M.G. Pia and W. Pokorski and J.M. Quesada and M. Raine and M.A. Reis and A. Ribon and A. {Ristić Fira} and F. Romano and G. Russo and G. Santin and T. Sasaki and D. Sawkey and J.I. Shin and I.I. Strakovsky and A. Taborda and S. Tanaka and B. Tomé and T. Toshito and H.N. Tran and P.R. Truscott and L. Urban and V. Uzhinsky and J.M. Verbeke and M. Verderi and B.L. Wendt and H. Wenzel and D.H. Wright and D.M. Wright and T. Yamashita and J. Yarba and H. Yoshida},
}

@article{PandoraPFA,
    title = {Particle flow calorimetry and the PandoraPFA algorithm},
    journal = {Nuclear Instruments and Methods in Physics Research Section A: Accelerators, Spectrometers, Detectors and Associated Equipment},
    volume = {611},
    number = {1},
    pages = {25-40},
    year = {2009},
    issn = {0168-9002},
    doi = {https://doi.org/10.1016/j.nima.2009.09.009},
    author = {M.A. Thomson},
}

@inproceedings{Arbor2014,
    author = "Ruan, Manqi and Videau, Henri",
    title = "{Arbor, a new approach of the Particle Flow Algorithm}",
    booktitle = "{International Conference on Calorimetry for the High Energy Frontier}",
    eprint = "1403.4784",
    archivePrefix = "arXiv",
    primaryClass = "physics.ins-det",
    reportNumber = "AIDA-CONF-2014-002",
    pages = "316--324",
    year = "2013"
}

@article{Arbor2018,
    author    = {Manqi Ruan and Hang Zhao and Gang Li and Chengdong Fu and Zhigang Wang and Xinchou Lou and Dan Yu and Vincent Boudry and Henri Videau and Vladislav Balagura and Jean-Claude Brient and Peizhu Lai and Chia-Ming Kuo and Bo Liu and Fenfen An and Chunhui Chen and Soeren Prell and Bo Li and Imad Laketineh},
    title     = {Reconstruction of physics objects at the Circular Electron Positron Collider with Arbor},
    journal   = {The European Physical Journal C},
    volume    = {78},
    number    = {5},
    pages     = {426},
    year      = {2018},
    publisher = {Springer},
    doi       = {10.1140/epjc/s10052-018-5876-z},
}

@article{HoughTransformation,
    title = {METHOD AND MEANS FOR RECOGNIZING COMPLEX PATTERNS},
    author = {Hough, P V.C.},
    doi = {},
    journal = {US Patent},
    number = {3},
    volume = {3,069,654},
    year = {1962},
    month = {12}
}

@article{HoughAlphaRho,
    title={Use of the Hough transformation to detect lines and curves in pictures},
    author={Richard O. Duda and Peter E. Hart},
    journal={Commun. ACM},
    year={1972},
    volume={15},
    pages={11-15},
}

@article{SDHCAL-Hough,
  title={{Tracking within Hadronic Showers in the CALICE SDHCAL prototype using a Hough Transform Technique}},
  author={Deng, Z and Li, Y and Wang, Y and Yue, Q and Yang, Z and Boumediene, D and Carloganu, C and Fran{\c{c}}ais, V and Cho, G and Kim, DW and others},
  journal={Journal of Instrumentation},
  volume={12},
  number={05},
  pages={P05009},
  year={2017},
  publisher={IOP Publishing},
}

@misc{CEPCSW,
  title        = {{CEPC offline software prototype based on Key4hep}},
  author       = {{The CEPC Study Group}},
  doi = {https://github.com/cepc/CEPCSW/},
  url = {https://github.com/cepc/CEPCSW/},
  year         = {2023},
}

@misc{CEPC-Reference-Detector-TDR,
      title={{CEPC Technical Design Report -- Reference Detector}}, 
      author={{The CEPC Study Group}},
      year={2025},
      eprint={2510.05260},
      archivePrefix={arXiv},
      primaryClass={hep-ex},
      url={https://arxiv.org/abs/2510.05260}, 
}

@article{Akopdzhanov:1976pr,
    author = "Akopdzhanov, G. A. and Inyakin, A. V. and Kachanov, V. A. and Krasnokutsky, R. N. and Lednev, A. A. and Mikhailov, Yu. V. and Prokoshkin, Yu. D. and Razuvaev, E. A. and Shuvalov, R. S.",
    title = "{Determination of Photon Coordinates in Hodoscope Cherenkov Spectrometer}",
    reportNumber = "IFVE-76-110",
    doi = "10.1016/0029-554X(77)90358-5",
    journal = "Nucl. Instrum. Meth.",
    volume = "140",
    pages = "441",
    year = "1977"
}

@article{Qi:2025dvo,
    author = "Qi, Baohua and Guo, Fangyi and Liu, Yong and Zhao, Zhiyu",
    title = "{Development of a novel high granularity crystal electromagnetic calorimeter}",
    doi = "10.1051/epjconf/202532000012",
    journal = "EPJ Web Conf.",
    volume = "320",
    pages = "00012",
    year = "2025"
}

@article{Yuda:1969xi,
    author = "Yuda, T.",
    title = "{Electron-induced cascade showers in inhomogeneous media}",
    doi = "10.1016/0029-554X(69)90401-7",
    journal = "Nucl. Instrum. Meth.",
    volume = "73",
    pages = "301--312",
    year = "1969"
}

@article{Amaldi:1980uz,
    author = "Amaldi, Ugo",
    title = "{Fluctuations in Calorimetry Measurements}",
    reportNumber = "CERN-EP-80-212",
    doi = "10.1088/0031-8949/23/4A/012",
    journal = "Phys. Scripta",
    volume = "23",
    pages = "409",
    year = "1981"
}

@article{SiW-2008,
doi = {10.1088/1748-0221/3/08/P08001},
url = {https://doi.org/10.1088/1748-0221/3/08/P08001},
year = {2008},
month = {aug},
publisher = {},
volume = {3},
number = {08},
pages = {P08001},
author = {{The CALICE collaboration} and J Repond and J Yu and C M Hawkes and Y Mikami and O Miller and N K Watson and J A Wilson and G Mavromanolakis and M A Thomson and D R Ward and W Yan and F Badaud and D Boumediene and C Cârloganu and R Cornat and P Gay and Ph Gris and S Manen and F Morisseau and L Royer and G C Blazey and D Chakraborty and A Dyshkant and K Francis and D Hedin and G Lima and V Zutshi and J-Y Hostachy and L Morin and E Garutti and V Korbel and F Sefkow and M Groll and G Kim and D-W Kim and K Lee and S Lee and K Kawagoe and Y Tamura and D A Bowerman and P D Dauncey and A-M Magnan and C Noronha and H Yilmaz and O Zorba and V Bartsch and J M Butterworth and M Postranecky and M Warren and M Wing and M Faucci Giannelli and M G Green and F Salvatore and T Wu and D Bailey and R J Barlow and M Kelly and S Snow and R J Thompson and M Danilov and V Kochetkov and N Baranova and P Ermolov and D Karmanov and M Korolev and M Merkin and A Voronin and B Bouquet and S Callier and F Dulucq and J Fleury and H Li and G Martin-Chassard and F Richard and Ch de la Taille and R Poeschl and L Raux and M Ruan and N Seguin-Moreau and F Wicek and Z Zhang and M Anduze and V Boudry and J-C Brient and C Clerc and G Gaycken and C Jauffret and A Karar and P Mora de Freitas and G Musat and M Reinhard and A Rougé and A L Sanchez and J-Ch Vanel and H Videau and J Zacek and J Cvach and P Gallus and M Havranek and M Janata and M Marcisovsky and I Polak and J Popule and L Tomasek and M Tomasek and P Ruzicka and P Sicho and J Smolik and V Vrba and J Zalesak and Yu Arestov and A Baird and R N Halsall and S W Nam and I H Park and J Yang},
title = {{Design and electronics commissioning of the physics prototype of a Si-W electromagnetic calorimeter for the International Linear Collider}},
journal = {Journal of Instrumentation},
}

@article{SiW-2009,
title = {{Response of the CALICE Si-W electromagnetic calorimeter physics prototype to electrons}},
journal = {Nuclear Instruments and Methods in Physics Research Section A: Accelerators, Spectrometers, Detectors and Associated Equipment},
volume = {608},
number = {3},
pages = {372-383},
year = {2009},
issn = {0168-9002},
doi = {https://doi.org/10.1016/j.nima.2009.07.026},
url = {https://www.sciencedirect.com/science/article/pii/S0168900209014673},
author = {C. Adloff and Y. Karyotakis and J. Repond and J. Yu and G. Eigen and C.M. Hawkes and Y. Mikami and O. Miller and N.K. Watson and J.A. Wilson and T. Goto and G. Mavromanolakis and M.A. Thomson and D.R. Ward and W. Yan and D. Benchekroun and A. Hoummada and M. Krim and M. Benyamna and D. Boumediene and N. Brun and C. Cârloganu and P. Gay and F. Morisseau and G.C. Blazey and D. Chakraborty and A. Dyshkant and K. Francis and D. Hedin and G. Lima and V. Zutshi and J.-Y. Hostachy and L. Morin and N. D’Ascenzo and U. Cornett and D. David and R. Fabbri and G. Falley and K. Gadow and E. Garutti and P. Göttlicher and T. Jung and S. Karstensen and V. Korbel and A.-I. Lucaci-Timoce and B. Lutz and N. Meyer and V. Morgunov and M. Reinecke and F. Sefkow and P. Smirnov and A. Vargas-Trevino and N. Wattimena and O. Wendt and N. Feege and M. Groll and J. Haller and R.-D. Heuer and S. Richter and J. Samson and A. Kaplan and H.-Ch. Schultz-Coulon and W. Shen and A. Tadday and B. Bilki and E. Norbeck and Y. Onel and E.J. Kim and N.I. Baek and D.-W. Kim and K. Lee and S.C. Lee and K. Kawagoe and Y. Tamura and D.A. Bowerman and P.D. Dauncey and A.-M. Magnan and H. Yilmaz and O. Zorba and V. Bartsch and M. Postranecky and M. Warren and M. Wing and M. {Faucci Giannelli} and M.G. Green and F. Salvatore and M. Bedjidian and R. Kieffer and I. Laktineh and D.S. Bailey and R.J. Barlow and M. Kelly and R.J. Thompson and M. Danilov and E. Tarkovsky and N. Baranova and D. Karmanov and M. Korolev and M. Merkin and A. Voronin and A. Frey and S. Lu and K. Prothmann and F. Simon and B. Bouquet and S. Callier and P. Cornebise and J. Fleury and H. Li and F. Richard and Ch. {de la Taille} and R. Poeschl and L. Raux and M. Ruan and N. Seguin-Moreau and F. Wicek and M. Anduze and V. Boudry and J.-C. Brient and G. Gaycken and P. {Mora e Freitas} and G. Musat and M. Reinhard and A. Rougé and J.-Ch. Vanel and H. Videau and K.-H. Park and J. Zacek and J. Cvach and P. Gallus and M. Havranek and M. Janata and M. Marcisovsky and I. Polak and J. Popule and L. Tomasek and M. Tomasek and P. Ruzicka and P. Sicho and J. Smolik and V. Vrba and J. Zalesak and B. Belhorma and M. Belmir and S.W. Nam and I.H. Park and J. Yang and J.-S. Chai and J.-T. Kim and G.-B. Kim and J. Kang and Y.-J. Kwon}
}

@Article{SiW-2022,
AUTHOR = {Pöschl, Roman},
TITLE = {{The CALICE SiW ECAL Technological Prototype—Status and Outlook}},
JOURNAL = {Instruments},
VOLUME = {6},
YEAR = {2022},
NUMBER = {4},
ARTICLE-NUMBER = {75},
URL = {https://www.mdpi.com/2410-390X/6/4/75},
ISSN = {2410-390X},
DOI = {10.3390/instruments6040075}
}

@misc{SiW-2020,
    title={{Analysis of SiW-ECAL technological prototype beam test with electron beam}},
    author={Yu Kato and Kiichi Goto and Taikan Suehara and {CALICE SiW-ECAL group}},
    year={2020},
    eprint={2002.12019},
    archivePrefix={arXiv},
    primaryClass={physics.ins-det}
}

\end{document}